\documentstyle[prl,aps,floats]{revtex}
\begin{document} 

\twocolumn[\hsize\textwidth\columnwidth\hsize\csname %
@twocolumnfalse\endcsname

\title{First principles theory of the EPR ${\rm g}$-tensor in 
solids: defects in quartz}

\author{Chris J. Pickard} 
\address{TCM Group, Cavendish Laboratory, Madingley Road,\\
Cambridge, CB3 0HE, United Kingdom}

\author{Francesco Mauri} 
\address{Laboratoire de Min\'{e}ralogie-Cristallographie de Paris,\\ 
Universit\'{e} Pierre et Marie Curie, 4 Place Jussieu, 75252, Paris, 
Cedex 05, France}

\date{\today}
\maketitle

\begin{abstract}
A theory for the reliable prediction of the EPR ${\rm g}$-tensor for
paramagnetic defects in solids is presented. It is based on density
functional theory and on the gauge including projector augmented wave
(GIPAW) approach to the calculation of all-electron magnetic response.
The method is validated by comparison with existing quantum chemical
and experimental data for a selection of diatomic radicals. We then
perform the first prediction of EPR ${\rm g}$-tensors in the solid
state and find the results to be in excellent agreement with experiment
for the $E'_1$ and substitutional P defect centers in quartz.
\end{abstract}
\vspace{0.2in}
]
\vspace{-0.575in}

\newcommand{\ket}    [1]{{|#1\rangle}}
\newcommand{\bra}    [1]{{\langle#1|}}
\newcommand{\braket} [2]{{\langle#1|#2\rangle}}
\newcommand{\bracket}[3]{{\langle#1|#2|#3\rangle}}


Electron paramagnetic resonance (EPR), also known as electron spin
resonance (ESR), is the most powerful spectroscopic technique for the
study of paramagnetic defects in solids. Indeed, defect centers are
often named directly after their EPR spectra. Applications of EPR
extend to any situation where there are unpaired electrons, including
the understanding of reactions involving free radicals in both
biological and chemical contexts or the study of the structure and
spin state of transition metal complexes.

EPR spectra of spin $1/2$ centers are made up of two contributions:
(i) the hyperfine parameters, which can be computed from the ground
state spin density, and have been used to connect theoretical studies
of defects to available experimental
data \cite{vandewalle,boero1997,blochl2000,pacchioni2001,uchino2001a,laegsgaard2000}, 
and (ii) the ${\rm
g}$-tensor.  Only recently have there been attempts to calculate the
${\rm g}$-tensor in molecules from first principles using density
functional theory
(DFT)\cite{schreckenbach97,malkina2000}.  However,
these approaches are valid only for finite systems and, thus, are not
useful for the calculation of the ${\rm g}$-tensor for paramagnetic
defects in solids, except possibly within a cluster approximation.  In
the absence of a predictive scheme, experimentally determined ${\rm
g}$-tensors are, of necessity, interpreted in terms of their symmetry
alone, leaving any remaining information unexploited.  A reliable,
first principles approach to the prediction of ${\rm g}$-tensors in
solids, in combination with structural and energetic calculations,
would access this information, and could be used for an unequivocal
discrimination between competing microscopic models proposed for
defect centers.  In this letter we describe an approach for the
calculation of the ${\rm g}$-tensor in extended systems, using
periodic boundary conditions and super-cells.

In a previous paper\cite{pickard2001-1} we have shown how to compute
the all-electron magnetic linear response, in finite and extended
systems, using DFT and pseudopotentials.  To achieve this we
introduced the gauge including projector augmented wave (GIPAW)
method, which is an extension of Bl\"ochl's projector augmented wave
(PAW) method\cite{blochl94}.  In Ref. \cite{pickard2001-1}, we used
GIPAW to compute the NMR chemical shifts in molecules and solids.
Here, we apply the GIPAW approach to the first principles prediction
of EPR ${\rm g}$-tensors for paramagnetic defects in solids.  We
validate our theory and implementation for diatomic radicals, for
which both all-electron quantum chemical calculations and experimental
data exist. As, until now, there have been no first principles
calculations of ${\rm g}$-tensors in solids, we validate our method in
the solid state by a direct comparison with experiment. In particular,
we interpret, from first principles, the EPR spectrum of the well
characterized and technologically important defects in quartz, the
$E'_1$ and P substitutional centers.



The ${\rm g}$-tensor is an experimentally defined quantity, arising
from the recognition that the EPR spectrum can be modeled using the
following effective Hamiltonian, bilinear in the total electron spin
${\bf S}$, and the applied uniform magnetic field or nuclear spins,
${\bf B}$ and ${\bf I}_{I}$, respectively:
\begin{equation}
\label{effectiveH}
H_{\rm eff} = \frac{\alpha}{2} {\bf S} \cdot {\bf g} \cdot {\bf B} +
\sum_{I} {\bf S} \cdot {\bf A}_{I} \cdot {\bf I}_{I}.
\end{equation}  
Here, and in the following, atomic units are used, $\alpha$ is the
fine structure constant, and the summation $I$ runs over the nuclei.
The tensors ${\bf A}_I$ are the hyperfine parameters (a PAW based
theory for its calculation has been described elsewhere by Van de
Walle and Bl\"ochl\cite{vandewalle,blochl2000}), and the tensor ${\bf
g}$ is the EPR ${\rm g}$-tensor.

In order to calculate the ${\rm g}$-tensor we start from the
electronic Hamiltonian which includes terms up to order $\alpha^3$, in
the presence of a constant external magnetic field ${\bf B}$
\cite{harriman1978,schreckenbach97}:
\begin{eqnarray}
\label{ab-initio}
H&=& \sum_i \left\{\frac{[{{\bf p}_i}+\alpha {\bf A}({\bf r}_i)]^2}{2}
- \sum_I \frac{Z_I}{\left|{\bf r}_i-{\bf R}_I\right|} +\sum_{j\ne
i}\frac{1}{\left|{\bf r}_i-{\bf r}_j\right|}\right\}\nonumber \\ & &
+H_{\rm Z} + H_{\rm Z-KE} + H_{\rm SO} + H_{\rm SOO}.
\end{eqnarray}
The summations over $i$ and $j$ run over the electrons and $H_{\rm
Z}$, $H_{\rm Z-KE}$, $H_{\rm SO}$, and $H_{\rm SOO}$ are the electron
Zeeman, the electron Zeeman kinetic energy correction, the spin-orbit,
and the spin-other-orbit terms respectively:
\begin{eqnarray}
&H&_{\rm Z} = \frac{\alpha {\rm g}_{\rm e}}{2} \sum_i{\bf
S}_i\cdot{\bf B} \nonumber\\ &H&_{\rm Z-KE} = -\frac{\alpha^3{\rm
g}_{\rm e}}{2}\sum_i \frac{p_i^2}{2} {\bf S}_i\cdot{\bf B}\nonumber\\
&H&_{\rm SO} = \frac{\alpha^2{\rm g}'}{4}\sum_i {\bf
S}_i\cdot\left(\sum_I Z_I\frac{{\bf r}_i-{\bf R}_I} {\left|{\bf
r}_i-{\bf R}_I\right|^3} -\sum_{j\ne i}\frac{{\bf r}_i-{\bf r}_j}
{\left|{\bf r}_i-{\bf r}_j\right|^3}\right)\nonumber\\ & & \times[{\bf
p}_i+\alpha{\bf A}({\bf r}_i)]\nonumber\\ &H&_{\rm SOO} = \alpha^2
\sum_{i,j\ne i}{\bf S}_i\cdot\frac{{\bf r}_i -{\bf r}_j}{|{\bf
r}_i-{\bf r}_j|^3}\times[{\bf p}_j+\alpha{\bf A}({\bf r}_j)].
\end{eqnarray}
The constant ${\rm g}'$ is related to $g_{\rm e}$, the electronic
Zeeman ${\rm g}$-factor in vacuum, by ${\rm g}'=2({\rm g}_{\rm e}-1)$,
and ${\bf A}({\bf r}) = \frac{1}{2}{\bf B}\times{\bf r}$ is the vector
potential.


Starting from the Hamiltonian of Eq. (\ref{ab-initio}), we can expand
the total energy in powers of $\alpha$, up to $O(\alpha^3)$, using
perturbation theory.  In the resulting expansion, the term bilinear in
${\bf S}_i$ and ${\bf B}$ is identified as the first term of
Eq. (\ref{effectiveH}).  This term can be rewritten within the
formalism of spin polarized DFT to obtain an explicit expression for
the g-tensor:
\begin{equation}
{\bf g} = {\bf g}_{\rm e} + \Delta{\bf g}_{\rm Z-KE} + \Delta{\bf
g}_{\rm SO} + \Delta{\bf g}_{\rm SOO} = {\bf g_{\rm e}} + \Delta{\bf
g},
\end{equation}
where ${\bf g}_{\rm e}={\rm g}_{\rm e}{\bf I}$, ${\bf I}$ being the
identity matrix, and:
\begin{equation}
\label{DgZ-KE}
\Delta {\bf g}_{\rm Z-KE} = -{\alpha}^2{\rm g}_{\rm e}
(T^{(0)}_{\uparrow}-T^{(0)}_{\downarrow}){\bf I}
\end{equation}
\begin{eqnarray}
\label{DgSO}
\Delta {\bf g}_{\rm SO}\cdot{\bf B} &=& \frac{\alpha}{2}{\rm g}{'}\int
d^3r [{\bf j}^{(1)}_{\uparrow}({\bf r})\times \nabla V^{(0)}_{\rm
ks,{\uparrow}}({\bf r})\nonumber \\ & &-{\bf
j}^{(1)}_{\downarrow}({\bf r}) \times \nabla V^{(0)}_{\rm
ks,{\downarrow}}({\bf r}) ]
\end{eqnarray}
\begin{equation}
\label{DgSOO}
\Delta {\bf g}_{\rm SOO}\cdot {\bf B}=2\int d^3r {\bf B}^{(1)}({\bf
r}) [\rho^{(0)}_{\uparrow}({\bf r})-\rho^{(0)}_{\downarrow}({\bf r})]
\end{equation}
Here $\uparrow$ denotes the majority spin channel and
$\rho^{(0)}_{\uparrow}({\bf r})$, $T^{(0)}_{\uparrow}$, and
${V}^{(0)}_{{\rm ks},\uparrow}({\bf r})$ are the unperturbed electron
probability-density, kinetic energy, and Kohn-Sham potential of the
$\uparrow$-spin channel, respectively. ${\bf j}^{(1)}_{\uparrow}({\bf
r})$ is the electronic charge-current linearly induced by the constant
magnetic field ${\bf B}$ in the $\uparrow$-spin channel. Finally,
${\bf B}^{(1)}({\bf r})$ is the magnetic field produced by the total
induced current, $[{\bf j}^{(1)}_{\uparrow}({\bf r})+{\bf
j}^{(1)}_{\downarrow}({\bf r})]$, which we correct for
self-interaction by removing the contribution from the current of the
unpaired electron, $[{\bf j}^{(1)}_{\uparrow}({\bf r})-{\bf
j}^{(1)}_{\downarrow}({\bf r})]$.


We can interpret the physical origin of deviation of the ${\rm
g}$-tensor from its value in vacuum.  The spin-other-orbit correction,
$\Delta {\bf g}_{\rm SOO}$, describes the screening of the external
field ${\bf B}$ by the induced electronic currents, as experienced by
the unpaired electron.  The unpaired electron itself is not at rest
and in the reference frame of the unpaired electron the electric field
due to the ions and to the other electrons is Lorentz transformed so
as to appear as a magnetic field. The interaction between the spin of
the unpaired electron and this magnetic field results in the the
spin-orbit correction, $\Delta {\bf g}_{\rm SO}$\cite{EPRfootnote}.
Finally, the electron Zeeman kinetic energy correction, $\Delta {\bf
g}_{\rm Z-KE}$, is a purely kinematic relativistic correction.


Eqs. (\ref{DgZ-KE}-\ref{DgSOO}) show that the evaluation of the
g-tensor requires, besides ground state quantities, the linear
magnetic response currents ${\bf j}^{(1)}({\bf r})$.  Mauri, Pfrommer
and Louie \cite{mauri96II} showed how to calculate the magnetic
response of a system of electrons in an infinite insulating crystal,
and our recent paper\cite{pickard2001-1} reformulated this so as to be
strictly valid for non-local pseudopotentials, and to reproduce the
valence all-electron currents even within the pseudisation core
region.  An accurate description of the all-electron currents in the
core regions is essential for the evaluation of the SO term,
Eq. (\ref{DgSO}).  Indeed, the dominant contribution to the integral
in Eq. (\ref{DgSO}) comes from the core region as a result of the
divergence of $V^{(0)}_{\rm ks}({\bf r})$ at the nuclei.


Using our GIPAW approach to the calculation of all-electron magnetic
response using pseudopotentials, described in detail in
Ref. \cite{pickard2001-1}, we break the SO term into three parts which
derive from the three GIPAW contributions to the induced current,
Eq. (34) of Ref. \cite{pickard2001-1}:
\begin{equation}
\Delta {\bf g}_{\rm SO} = \Delta {\bf g}_{\rm SO}^{\rm bare} + \Delta
{\bf g}_{\rm SO}^{\rm \Delta d} + \Delta {\bf g}_{\rm SO}^{\rm \Delta
p}.
\end{equation}
The $\Delta {\bf g}_{\rm SO}^{\rm bare}$ term is evaluated from
Eq. (\ref{DgSO}) using a spin dependent version of the ${\bf
j}^{(1)}_{\rm bare}({\bf r})$ of Ref. \cite{pickard2001-1} and a local
Kohn-Sham potential consisting of a sum of the self-consistent
contribution to the local potential and the local parts of the
pseudopotentials.

The diamagnetic correction term $\Delta {\bf g}_{\rm SO}^{\rm \Delta
d}$ can be evaluated from the ground-state valence pseudo wavefunctions
$| \bar \Psi_{o,\uparrow}^{(0)}\rangle$ using the following expression
:
\begin{eqnarray}
\Delta {\bf g}_{\rm SO}^{\rm \Delta d}\cdot{\bf B}=&&
\sum_{{I},o,n,m}\langle \bar \Psi_{o,\uparrow}^{(0)}| \tilde
p_{{I},n}\rangle {\bf e}^{I}_{n,m} \langle \tilde p_{{I},m}| \bar
\Psi_{o,\uparrow}^{(0)}\rangle\nonumber\\ &&- \sum_{{I},o,n,m}\langle
\bar \Psi_{o,\downarrow}^{(0)}| \tilde p_{{I},n}\rangle {\bf
e}^{I}_{n,m} \langle \tilde p_{{I},m}| \bar
\Psi_{o,\downarrow}^{(0)}\rangle.
\end{eqnarray}
The summation $o$ is over occupied states. 
The projector functions $|\tilde p_{{I},n}\rangle$ are defined
in Ref. \cite{pickard2001-1} and satisfy $\langle\tilde
p_{{I},n}|\tilde\phi_{{I}',m}\rangle= \delta_{{I},{ I}'}
\delta_{n,m}$, where $|\tilde\phi_{{I},n}\rangle$ are a set of
pseudo-partial-waves corresponding to the all-electron partial waves
$|\phi_{{I},n}\rangle$. The projector weights ${\bf e}^{I}_{n,m}$ are
given by the following atom centered integrals:
\begin{eqnarray}
\label{e}
{\bf e}^{I}_{n,m}=-\frac{\alpha^2{\rm g}'}{4}&&[ \langle \phi_{{I},n}|
({\bf B}\times{\bf r})\times\nabla V({r})
|\phi_{{I},m}\rangle\nonumber\\ &&- \langle \tilde \phi_{{I},n}| ({\bf
B}\times{\bf r})\times\nabla\tilde V({r}) |\tilde\phi_{{I},m}\rangle]
\end{eqnarray}
The potentials $V(r)$ and $\tilde V(r)$ in Eqs. (\ref{e})
and (\ref{f}) are the screened atomic all-electron and local channel
pseudopotentials respectively.

The evaluation of the paramagnetic correction term $\Delta {\bf
g}_{\rm SO}^{\rm \Delta p}$ is more involved as it requires the first
order linear response wavefunctions. However, the required evaluation
can be described by analogy with the calculation of paramagnetic
correction to the NMR chemical shifts, $\sigma_{\rm GIPAW}^{\Delta {\rm
p}}$, replacing the weights ${\bf f}^{I}_{n,m}$ in Eq. (60) of
Ref. \cite{pickard2001-1} by
\begin{equation}
\label{f}
{\bf f}^{I}_{n,m}= \langle \phi_{{I},n}| \frac{{\rm g}'}{2}\frac{{\bf
L}}{r}\frac{\partial V(r)}{\partial r} |\phi_{{I},m}\rangle -
\langle\tilde\phi_{{I},n}| \frac{g'}{2}\frac{{\bf
L}}{r}\frac{\partial\tilde V(r)}{\partial r}
|\tilde\phi_{{I},m}\rangle,
\end{equation}
where ${\bf L}$ is the angular momentum operator.

The electron Zeeman kinetic energy correction term $\Delta {\bf
g}_{\rm Z-KE}$ is evaluated by combining a straightforward PAW
correction with the quantity evaluated from the ground-state pseudo
valence wavefunctions using Eq. (\ref{DgZ-KE}). In this work the SOO
term is evaluated from the induced field ${\bf B}^{(1)}({\bf r})$
derived from the bare induced current, and the spin density due to the
pseudo wavefunctions. It is expected that a full GIPAW treatment would
result in only minor corrections since (i) the SOO term is small in
comparison to the SO term, and (ii) both the induced field and the spin
density do not diverge at the nuclei.
 

To validate our new expressions for the evaluation of the ${\rm
g}$-tensor, and our implementation of them into a parallelized
plane-wave pseudopotential code, we compare with the all-electron
gauge including atomic orbital (GIAO) DFT results obtained by
Schreckenbach and Ziegler\cite{schreckenbach97} for a series of
diatomic radicals. We use their calculated bond lengths for the
dimers, but approximate the isolated dimers by using large
super-cells.  Troullier-Martins pseudopotentials\cite{tm:vps} and the
(spin polarized) generalized gradient approximation due to Perdew
\emph{et al}\cite{perdew96} (GGA-PBE) are used throughout our
calculations. Table \ref{DIATOMICSvsSZ} shows the excellent agreement
between our two approaches. The exception is the AlO radical, for
which we obtain much closer agreement with experiment. The otherwise
close agreement between these two very different approaches suggests a
technical rather than fundamental problem in the GIAO calculation for
AlO. Comparison with experiment is made through Table 2 of
Ref. \cite{schreckenbach97}, while acknowledging that most
measurements are performed in solid matrices, which strongly influence
the ${\rm g}$-tensor (most notably the $\Delta {\rm g}_\parallel$
components), and that the experimental errors are of the order of
several hundred parts per million (ppm).

Finally, by analyzing the different contributions, including the SOO
term, we found that in all dimers apart from H$_2^+$, the SO term
accounts for more than 90\% of $Tr \Delta {\rm g}/3$, and that the
paramagnetic correction term $\Delta {\bf g}_{\rm SO}^{\rm \Delta p}$
accounts for the overwhelming majority of the SO term.


To further validate our approach to the calculation of the ${\rm
g}$-tensor and to apply it for the first time in the solid state, we
study two defects of $\alpha$-quartz.

The $E'_1$ center is associated with a positively charged oxygen
vacancy, with the unpaired electron on a Si dangling bond.  As in
previous calculations \cite{boero1997,blochl2000,carbonaro2001}, we
model the defect with a 71 atom (24 Si and 47 O), positively charged
(+1) hexagonal super-cell.  We use the theoretical GGA-PBE lattice
parameters (which are 1\% larger than in experiment) and relax the
atoms. For the structural optimization we use a $\Gamma$ only k-point
sampling and a plane-wave cutoff of 50 Ry.  The resulting relaxed
structure is very close to that of Ref. \cite{blochl2000}.
The EPR ${\rm g}$-tensor is calculated using our relaxed structure, a
plane-wave cutoff of 70 Ry and 4 inequivalent k-points.  In Table
\ref{EP1} we compare our theoretical ${\rm g}$-tensor with the
experimental results\cite{jani83}, finding excellent agreement.

The P2 defect center is neutral and assocated to a four-fold 
coordinated P atom, substituting for a Si atom.
The center exists as two variants at low
temperature ($<$140K) in quartz, labelled P2(I) in the ground state
and P2(II) in the excited state \cite{uchida1979}.  
Only recently have these P defect centers
been examined using DFT based total energy approaches
\cite{laegsgaard2000,pacchioni2001}.  However, up until
now, the connection with experiment has been made using the
eigenvalues of the hyperfine parameters alone, and the two variants of
the defects, P2(I) and P2(II), have not been distinguished
theoretically.

Using the method described above, and a super-cell of 72 atoms, 5
distinct total energy local minima are found as a function of the
initial configuration, Table \ref{P2-energetics}.  The configuration
with the highest energy corresponds to a symmetric relaxation with the
P remaining tetrahedrally coordinated, and O-P-O angles of about
109$^\circ$. In the 4 other configurations the P atom moves off-center,
opening up one of the 6 O-P-O angles, which reaches a value of about
150$^\circ$.  We computed the EPR ${\rm g}$-tensors for our two lowest
energy structures and compare them with the experimental
results\cite{uchida1979} in Table \ref{P2}.  Again, we obtain an
excellent agreement between theory and experiment, which confirms that
the two lowest energy theoretical structures correspond to the two
lowest energy experimental structures.  However, the comparison
between ${\rm g}$-tensors shows that the energy ordering between the
P2(I) and P2(II) species is not correctly described by theory.  This
is not surprising given the small energy separation between the two
configurations. This is expected to be sensitive to both the size of
the super-cell and the use of approximated DFT functionals.


To summarize, we have calculated the EPR ${\rm g}$-tensor for a
paramagnetic defect in an extended solid for the first time and find
our results to be in excellent agreement with experimental results for
the $E'_1$ defect. On applying the method to the P2 defect, we show
that comparison with experimental ${\rm g}$-tensors can provide
structural information where the accuracy of DFT for energetics is
insufficient. Combined with the calculation of hyperfine
parameters\cite{vandewalle,boero1997,blochl2000,pacchioni2001,uchino2001a,laegsgaard2000},
we expect that our GIPAW based first principles approach to the
prediction of EPR ${\rm g}$-tensors will be of great use in the
assessment of models proposed for less well characterized paramagnetic
defects, and add significantly to the tools available to the
electronic structure community.


C.J.P. would like to thank the Universit\'e Paris 6 and the
Universit\'e Paris 7 for support during his stay in Paris. The
calculations were performed at the IDRIS super-computing center of the
CNRS and on Hodgkin (SGI Origin 2000) at the University of Cambridge's
High Performance Computing Facility.


\begin{table}
\caption{Calculated $\Delta {\rm g}$ tensors, in parts per million
(ppm), for diatomic molecules. For comparison with
Ref. \protect\cite{schreckenbach97} we omit (in this table only) the
SOO contribution to our calculations. A 100 Ry plane-wave cutoff is
used.}
\label{DIATOMICSvsSZ}
\begin{tabular}{lrrrr}
Molecule & \multicolumn{2}{c}{$\Delta {\rm g}_\parallel $}
&\multicolumn{2}{c}{$\Delta {\rm g}_\perp$}  \\
         &  GIPAW  &     SZ\protect\cite{schreckenbach97}    
&  GIPAW  &     SZ\protect\cite{schreckenbach97}    \\\hline
H$_2^+$  &   -39   &    -39    &    -41  &    -42    \\
CO$^+$   &  -134   &   -138    &  -3223  &  -3129    \\
CN       &  -138   &   -137    &  -2577  &  -2514    \\
AlO      &  -141   &   -142    &  -2310  &   -222    \\
BO       &   -69   &    -72    &  -2363  &  -2298    \\
BS       &   -80   &    -83    &  -9901  &  -9974    \\
MgF      &   -49   &    -60    &  -2093  &  -2178    \\
KrF      &  -340   &   -335    &  61676  &  60578    \\
XeF      &  -333   &   -340    & 157128  & 151518      
\end{tabular}
\end{table}

\begin{table}
\caption{Calculated $\Delta {\bf g}$ tensors for our model $E'_1$
defect, and corresponding experimental data\protect\cite{jani83}.}
\label{EP1}
\begin{tabular}{cccccc}
\multicolumn{2}{c}{Principal values}&
\multicolumn{4}{c}{Principal directions}\\
GIPAW (ppm) & Expt. (ppm) & \multicolumn{2}{c}{GIPAW}   &
 \multicolumn{2}{c}{Expt.}    \\
            &             &  $\theta$    &   $\phi$     &
    $\theta$  & $\phi$        \\\hline
 -651       &   -530      & 110.0$^\circ$& 223.5$^\circ$&
 114.5$^\circ$& 227.7$^\circ$ \\
-2255       &  -1790      & 142.3$^\circ$& 341.6$^\circ$&
 134.5$^\circ$& 344.4$^\circ$ \\
-2481       &  -2020      & 120.4$^\circ$& 121.1$^\circ$&
 125.4$^\circ$& 118.7$^\circ$
\end{tabular}
\end{table}

\begin{table}
\caption{Calculated total energies, with respect to
our lowest energy configuration. For non-tetrahedral configurations,
OPO indicates the largest O-P-O angle
after the relaxation, and OSiO the corresponding angle in the
unrelaxed $\alpha$-quartz structure. With l and s we specify whether the 
two SiO bonds forming in the OSiO angle are short or long in the unrelaxed
structure. The number of symmetry equivalent structures is 
N$_{\rm c}$. We report our assignment of the experimental 
centers based on the comparison of Table \protect\ref{P2}. }
\label{P2-energetics}
\begin{tabular}{lcccccc}
SiO & OSiO& OPO & N$_{\rm c}$  & Energy (meV)& species \\\hline
\multicolumn{3}{c}{tetrahedral configuration}   &     1       & 730 &           \\
l s& 108.8$^\circ$   & 147.4$^\circ$  &     2       & 243 &           \\
s s& 108.9$^\circ$   & 155.4$^\circ$  &     1       & 178 &           \\
l l& 109.5$^\circ$   & 157.6$^\circ$  &     1       & 79 & P2(I)     \\
l s& 110.4$^\circ$   & 155.1$^\circ$  &     2       &  0 & P2(II) 
\end{tabular}
\end{table}

\begin{table}
\caption{Calculated $\Delta {\bf g}$ tensors for our model P2 defect
centers, and corresponding experimental data \protect\cite{uchida1979}. }
\label{P2}
\begin{tabular}{cccccc}
\multicolumn{2}{c}{Principal values}&
\multicolumn{4}{c}{Principal directions}\\
GIPAW (ppm) & Expt. (ppm) & \multicolumn{2}{c}{GIPAW}   &
 \multicolumn{2}{c}{Expt.}    \\
       &                  &  $\theta$    &   $\phi$     &
    $\theta$  & $\phi$        \\\hline
\multicolumn{6}{c}{Configuration P2(I)}\\

 1249       &    900      &  65.0$^\circ$&  90.0$^\circ$&
 64.8 $^\circ$&  90.0$^\circ$ \\
-980        &  -1100      &  90.0$^\circ$&   0.0$^\circ$&
  90.0$^\circ$&   0.0$^\circ$ \\
-3414       &  -3200      &  25.0$^\circ$& 270.0$^\circ$&
  25.2$^\circ$& 270.0$^\circ$\\
\multicolumn{6}{c}{Configuration P2(II)}\\
 1146       &   1100      &  47.1$^\circ$&  13.1$^\circ$&
 46.5 $^\circ$&  11.8$^\circ$ \\
-824        &  -1000      &  99.0$^\circ$& 94.7$^\circ$&
 101.1$^\circ$&  91.1$^\circ$ \\
-3454       &  -3200      & 135.7$^\circ$& 355.4$^\circ$&
 134.4$^\circ$& 350.1$^\circ$
\end{tabular}
\end{table}


\end{document}